\documentclass[prl,twocolumn,showpacs]{revtex4}
\usepackage{amsmath,amssymb,graphicx,bezier}
\usepackage{comment}
\newcommand{\nc}{\newcommand}
\nc{\rnc}{\renewcommand}
\nc{\nn}{\nonumber}
\nc{\ch}{\cosh}
\nc{\sh}{\sinh}
\nc{\sech}{{\rm sech}}
\nc{\bra}{\langle}
\nc{\ket}{\rangle}
\rnc{\H}{\mathcal{H}}
\nc{\M}{\mathcal{M}}
\rnc{\(}{\left(}
\rnc{\)}{\right)}
\rnc{\Im}{{\rm{Im}\,}}
\rnc{\Re}{{\rm{Re}\,}}
\nc{\tr}{{\rm Tr}}
\nc{\Lam}{\Lambda}
\nc{\ep}{\varepsilon}
\nc{\calH}{{\cal H}}
\nc{\calA}{{\cal A}}
\nc{\calL}{{\cal L}}
\def\bm#1{\mbox{\boldmath $#1$}}
\begin{document}
%
\title
{
Exact Analysis of ESR Shift in the Spin-1/2 Heisenberg 
Antiferromagnetic Chain
}
\author{Yoshitaka Maeda}
\thanks{Present Address:
Department of Physics and Mathematics,
Aoyama Gakuin University, Sagamihara,
Kanagawa 229-8558, Japan.}

\affiliation{ Department of Physics, Tokyo Institute of Technology \\
Oh-okayama, Meguro-ku, Tokyo 152-8551, Japan}

\author{Kazumitsu Sakai}

\affiliation{ Department of Physics, Tokyo Institute of Technology \\
Oh-okayama, Meguro-ku, Tokyo 152-8551, Japan}

\author{Masaki Oshikawa}
\affiliation{ Department of Physics, Tokyo Institute of Technology \\
Oh-okayama, Meguro-ku, Tokyo 152-8551, Japan}
%
%
\begin{abstract}
A systematic perturbation theory is developed for the ESR 
shift and is applied to the spin-1/2 Heisenberg antiferromagnetic chain
with a general anisotropic exchange interaction.
Using the Bethe ansatz technique, the resonance shift 
is obtained exactly for the whole range of temperature and magnetic
field, in the first order of the anisotropy.
The obtained $g$-shift strongly depends on magnetic fields
at low temperature, showing a significant deviation from
the previous classical result.
\end{abstract}
\maketitle
%
%
%
Many of the interesting phenomena and experimental measurements
in strongly correlated quantum systems are related to {\it dynamics} 
of the system. 
Despite the considerable progress in many-body theory,
available exact results on quantum dynamics in many-body systems
are still rather scarce.
Indeed, even a systematic framework for approximate
calculations is often not well established.
Electron spin resonance (ESR) is a typical example of such
dynamical phenomena~\cite{Ajiro}.

In ESR, 
a static magnetic field $H$ and an electromagnetic 
wave with the polarization  perpendicular to $H$ are applied to a 
magnetic system, and then the intensity of the radiation absorption 
$I(\omega)$ is measured as a function of frequency $\omega$ of the
electromagnetic wave.
%
Within the linear response theory, $I(\omega)$ is 
written as
$I(\omega)\propto\omega \chi^{\prime\prime}_{\sigma\sigma}(\omega)$, 
where  $\chi_{\sigma\sigma}^{\prime\prime}(\omega)$ is the imaginary part of 
the dynamical susceptibility $\chi_{\sigma\sigma}(\omega)$
{\em at zero momentum}:
\begin{equation}
\chi_{\sigma\sigma}(\omega)= \frac{i}{N}\int_0^{\infty}
       \bra [S^\sigma(t),S^\sigma(0)]\ket e^{i\omega t} \; dt.
\end{equation}
Here $N$ is the total number of spins, $S^\sigma$ is the total spin operator 
($S^\sigma=\sum_{j=1}^N S_j^\sigma$) and $\sigma$
is the direction of the polarization.
We take the $z$-axis as the direction of the magnetic field $H$, and
for simplicity we assume $\sigma \in\{x,y\}$.
Let us consider a system with Hamiltonian 
%
\begin{equation}\label{TotalH}
\H=\H_{SU(2)}+\H_Z+\H^{\prime}, 
\end{equation}
where  $\H_{SU(2)}$ is an $SU(2)$-symmetric term and $\H_Z=-HS^z$ is the 
Zeeman term, and $\H^{\prime}$ is an anisotropic term assumed to be a 
small perturbation. In the absence of $\H^{\prime}$, the energy absorption 
occurs only at the paramagnetic resonance frequency $\omega=H$.
It is only the anisotropic perturbation $\H^{\prime}$,
although it may be small, that causes the change in the spectrum
such as a shift and a broadening of the resonance peak. 

Several frameworks of perturbation theory in $\H^{\prime}$,
typified by the Kubo-Tomita theory~\cite{KT}, have been developed.
However, the questionable assumptions and technical difficulties
limited their success in systems with strong quantum fluctuations.
More recently, field theoretical~\cite{OA} and  numerical~\cite{MYO} 
approaches avoiding ambiguous assumptions have been 
provided and brought about a renewed interest in ESR theory.  
Unfortunately, however, the former is applicable to 1D systems 
only in low-energy regimes, and the latter is applied only 
to small systems.
Hence rather much remains open in ESR theory.

This is evident in theory on the shift
in the Heisenberg antiferromagnetic chain,
which is one of the most fundamental problems in ESR.
The ESR shift $\delta \omega$ is usually defined as the deviation
of the resonance frequency from the paramagnetic resonance frequency $\omega=H$.
Nagata-Tazuke theory~\cite{NT} published more than 30 years ago
has been the only theory for arbitrary temperature,
and still remains the standard. However, it is based on the classical spin
approximation and does not incorporate the quantum fluctuation.
In this letter,  we utilize the integrability of
the the spin-$1/2$ Heisenberg antiferromagnetic chain
to go beyond the limitations of the previous approaches.
As a result, we obtain the ESR shift for general anisotropic
interactions at arbitrary temperature and magnetic field, fully
including the quantum effects. 

%
Nagata-Tazuke theory~\cite{NT} is based on
the fundamental formula for the ESR shift, first proposed
in ref.~\cite{KaTa}.
Its simplest form reads:
\begin{equation}
\label{eq:formula}
\delta\omega =
-\frac{\langle [\calA,S^-]\rangle}{2\langle S^z\rangle},
\end{equation}
where $\langle\cdots\rangle$ is the thermal expectation value with
respect to $\H$,
and $\calA \equiv [\H^{\prime},S^+]$. 
In ref.~\cite{NT}, this formula is further evaluated for the
1D {\em classical} Heisenberg antiferromagnet in the
lowest order of $H$, 
utilizing Fisher's solution~\cite{Fisher} of the classical
Heisenberg chain.
Owing to the reliance on the classical limit, we cannot expect
their result to apply to the strongly quantum system.
On the other hand, as some nontrivial assumptions are involved
in the original derivation of the formula~(\ref{eq:formula}), 
in literatures there seems to be a confusion over the validity of
the starting point, eq.~(\ref{eq:formula}) itself.

Here, we clarify the validity and limitation of the
formula~(\ref{eq:formula}).
To discuss the resonance shift unambiguously,
in this letter, we define  the shift 
$\delta \omega_{\sigma\sigma}$ for the polarization $\sigma$
in terms of
the dynamical susceptibility as 
%
\begin{align}\label{eq:defshift}
\delta\omega_{\sigma\sigma}=
\frac{{\int_0^\infty d\omega \; \omega 
                       \chi_{\sigma\sigma}''(\omega)}}
                       {{\int_0^\infty d\omega  \;
                       \chi_{\sigma\sigma}''(\omega)}}-H.
\end{align}
Note that the integration interval is taken to be $[0,\infty]$ instead of 
$[-\infty,\infty]$, because
$\chi''_{\sigma\sigma}(\omega)$ is an odd function of
$\omega$.
By using  a sum rule~\cite{Kubobook}, the numerator of 
eq.~\eqref{eq:defshift} is explicitly written  as 
$-\pi\bra [[\H,S^\sigma],S^\sigma] \ket/2$.
In contrast, the integration in the 
denominator cannot be performed analytically.  Hence we shall 
evaluate~\eqref{eq:defshift} perturbatively.

The computation in the first order can be carried out without
resorting to any other approximation, by utilizing the exact
Heisenberg equation of motions such as
$\dot{S}^+=-i H S^+ + i \mathcal{A}$~\cite{OA}.
After straightforward but somewhat tedious calculations, we 
obtain the first order perturbative contribution to
eq.~\eqref{eq:defshift}, exactly as in eq.~(\ref{eq:formula}).
Namely, eq.~(\ref{eq:formula}) is generally {\em exact} in the
first order of the anisotropic perturbation~$\H'$.
Note that, however, eq.~(\ref{eq:formula}) breaks down in
the second and higher orders in $\H'$~\cite{MO,OA}.
Restricting eq.~(\ref{eq:formula}) to the first order in $\H'$,
we obtain the simplified expression
\begin{equation}
\label{eq:formula2}
\delta\omega_{\sigma\sigma}=
-\frac{\langle [\calA,S^-]\rangle_0}{2\langle S^z\rangle_0}
+O(\calH'^2),
\end{equation}
where $\langle\cdots\rangle_0$ is the thermal expectation value
with respect to the unperturbed term $\H_0 = \H_{SU(2)} + \H_Z$.

As there is no contribution to
(\ref{eq:formula2}) from satellite peaks~\cite{KT},
it exactly gives the shift
of the main paramagnetic resonance in the first order.
As we will see, for the exchange anisotropy (e.g. dipolar
interaction), the first order contribution is
generally
non-vanishing.
Thus the precise evaluation of eq.~(\ref{eq:formula2}) is
of  great importance.

Now, let us 
consider the Heisenberg chain
\begin{equation}
\H_0=
\calH_{SU(2)}+\calH_Z=J\sum_{j=1}^{N}
\bm S_j \cdot\bm S_{j+1}-H\sum_{j=1}^{N} S_j^z ,
\label{eq:Hchain}
\end{equation}
with a small anisotropic
interaction as a perturbation $\calH'$.
%
%
As $\calH'$, we consider the anisotropic exchange
interaction.
Since the antisymmetric exchange interaction (Dzyaloshinskii-Moriya
interaction) does not contribute in the first order, we only consider
the symmetric (with respect to an exchange
of neighboring spins) exchange interaction. The dipolar interaction
may also be approximated by a suitable exchange anisotropy in the
nearest neighbor interaction.
Assuming the chain is uniform, the anisotropic exchange interaction
can always be diagonalized as
\begin{equation}
\H'=\sum_j \sum_{p,q \in \{a,b,c\}} J'_{pq}S_j^pS_{j+1}^q,\quad
J'_{pq}=J'_p \delta_{pq},
\label{eq:H'} 
\end{equation}
where $a,b,c$ are the principal axes of the anisotropy.
Let the direction of the static magnetic field $H$ be
$(\alpha,\beta,\gamma)= (\sin\theta\cos\phi,\sin\theta\sin\phi,\cos\theta)$
in the $(a,b,c)$-coordinate,
where $\theta$ is zenith angle and $\phi$ is azimuth angle.
As introduced previously, the $(x,y,z)$-coordinate is defined so
that $H$ is applied in the $z$-axis.
The resonance shift~\eqref{eq:formula2} is now evaluated as
\begin{equation}
\delta\omega=f(\theta,\phi)\times Y(T,H).
\label{eq:shift}
\end{equation}
Here the direction-dependent factor $f(\theta,\phi)$ and the
field- and temperature-dependent factor 
$Y(T,H)$ are respectively given by 
\begin{align}
& f(\theta,\phi)=J'_a(1-3\alpha^2)+J'_b(1-3\beta^2)+J'_c(1-3\gamma^2),
\label{eq:f}
\\
&Y(T,H) = \frac{\langle S^z_jS^z_{j+1}-S^x_jS^x_{j+1}\rangle_0}{\langle
              S^z_j\rangle_0}.
\label{R}
\end{align}
In general, eqs.~\eqref{eq:shift}--\eqref{R} are valid for 
arbitrary spin, when we consider 
$\calH'$~\eqref{eq:H'} as a perturbation.
It would be interesting to evaluate them numerically, for example
by a Quantum Monte Carlo simulation.
However, this could be a challenging problem especially in
the weak field regime; there will be a cancellation of
significant digits in taking the difference between the
very close quantities $\langle S^z_j S^z_{j+1} \rangle$
and $\langle S^x_j S^x_{j+1} \rangle$.

Hereafter we shall concentrate on the $S=1/2$ case.
The total Hamiltonian~\eqref{TotalH} 
is {\it no longer} integrable even with $S=1/2$, due to the existence of 
the anisotropy~\eqref{eq:H'} and magnetic fields. 
The first-order shift~\eqref{R}, however, can be exactly 
calculated  by the Bethe ansatz 
technique~\cite{QTM} utilizing the integrability of the
zeroth order Hamiltonian~\eqref{eq:Hchain}.
%
%
%
After long calculations,
we obtain
\begin{equation}
Y(T,H)=\frac12-\frac{T}{2\pi J}\oint_{\Gamma} \ln(1+\eta(x+i)) \; dx,  
\label{NLIE}
\end{equation}
where the contour $\Gamma$ encloses the real axis counterclockwise:
for instance $\Gamma$ can be
the rectangular contour whose corners are $-\infty-i,+\infty-i,+\infty+i$,
and $-\infty+i$.

The unknown 
function $\eta(x)$ is determined by the following non-linear 
integral equation:
\begin{align}
\ln\eta(x)&=\frac{2\pi J}{T} a_1(x)
-\frac{H}{T}\nonumber\\
&-\oint_{\Gamma} a_2(x-y-i)\ln(1+\eta(y+i)) \; dy,
\label{eq:eta}
\end{align}
where $a_n(x)=n/(\pi(x^2+n^2))$.
%

Although the result~\eqref{NLIE} is limited 
only to the $S=1/2$ Heisenberg antiferromagnetic chain,
it gives the resonance 
shift \eqref{R} without the difficulty due to the cancellation of
significant digits.
As a consequence, the resonance shift can be evaluated
numerically with a very high
accuracy for arbitrary fields and temperatures.

In experimental studies, it is customary to discuss the shift in
terms of the effective $g$-factor
\begin{equation}
g_{\rm eff} \equiv g_{\infty}\biggl( 1 + \frac{\delta\omega}{H}\biggr) =
 g_{\infty} \biggl(1 + \frac{f(\theta,\phi)Y(T,H)}{H} \biggr),
\end{equation}
and the $g$-shift $\Delta g \equiv g_{\rm eff}-g_{\infty}$.
Here $\delta \omega$ is the frequency shift we have obtained
with setting $\mu_B g_{\infty} =1$, and
$g_{\infty}$ is the $g$-factor appearing
in the Zeeman term.
When $J'_p \ll J, H$, we can identify
$g_\infty$ with the observed $g$-factor
in the high temperature limit.

\begin{figure}[htb]
\begin{center}
\includegraphics[width=0.45\textwidth]{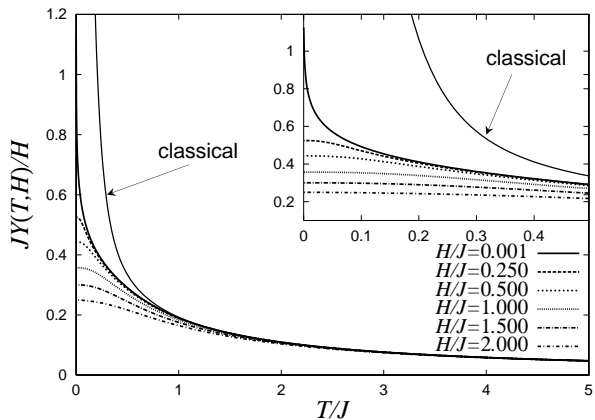}
\end{center}
\caption{The temperature dependence of $JY(T,H)/H$ for 
various $H$. For comparison, the result from the
classical approximation is also depicted.
Inset: details of the low-temperature.
}
\label{fig:QTMvsNT}
\end{figure}

In Fig.~\ref{fig:QTMvsNT}, the temperature dependence of $J Y(T,H) /H$,
which is proportional to the $g$-shift $\Delta g$,
is depicted for various magnetic fields.
We find that $\Delta g$ clearly depends on the magnetic fields $H$
especially in the low-temperature regime, and converges to a certain 
value at $T=0$ for any finite $H$. One also finds that $\Delta g$ decreases
with increasing $H$.
In the limit $H\to\infty$, $\Delta g$ vanishes at any temperature.
In contrast, for a weak magnetic field $H\ll 1$, $\Delta g$
logarithmically increases with decreasing temperature and approaches
infinity when $H\to 0$.

In Fig.~\ref{fig:QTMvsNT}, the present results
are compared to the previous classical result~\cite{NT}.
It was obtained by expanding eq.~\eqref{R} in the lowest order of $H$
and then applying Fisher's exact solutions to the classical Heisenberg
chain at $H=0$~\cite{Fisher}, as
\begin{align}
&Y_{\rm cl}(T,H)=
   -\frac{H K}{5 J}\left\{\frac{2-u(K)/K}{1-u(K)^2}+
    \frac{2K}3\right\}  ,              \nonumber\\
&u(K)=\coth K -\frac1K, \quad  K=-\frac{JS(S+1)}{T},
\label{NT}
\end{align}
where we shall set $S=1/2$ in the present case.
Obviously 
the $g$-factor for the classical approach
diverges with decreasing $T$, which contrasts to our results.
In addition, the classical result does not describe the
$H$-dependence as it is obtained in the lowest order expansion in $H$.

%
%
%
%
\begin{figure}[htb]
\begin{center}
\includegraphics[width=0.45\textwidth]{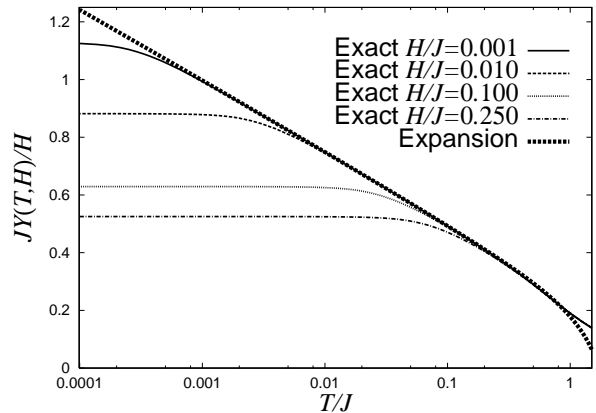}
\end{center}
\caption{Comparison of the exact result with the asymptotic 
behavior \eqref{eq:lowHT}
(we set $C=3/4$). A crossover from a 
regime dominated by logarithmic corrections
to a regime dominated  by algebraic corrections 
 occurs at $T\sim H$.}
\label{fig:lowT}
\end{figure}
%

Comparing the integral equations~(\ref{NLIE})~and~(\ref{eq:eta}) for $Y(T,H)$
and that for the magnetization $\langle S^z \rangle_0$~\cite{QTM,FKEG}
in the $S=1/2$ Heisenberg chain~(\ref{eq:Hchain}),
we find a remarkable identity relating them:
\begin{equation}\label{eq:S=M}
Y(T,H)=\frac1J\int \langle S^z_j\rangle_0 \; dJ .
\end{equation}
This is useful in obtaining the
low-temperature asymptotics of the shift.
Because the low-temperature asymptotics of $\bra S_j^z \ket_0$ 
is derived by an effective field theory~\cite{Lukyanov},
we obtain, for $H \ll T \ll J$,
\begin{align}\label{eq:lowHT}
  Y(T,H)=\frac{H}{J\pi^2}
           &\Bigl\{
               \calL+\frac12\ln\calL+\frac1{4\calL}+
               \frac{\ln\calL}{4\calL}-\frac{\sqrt{3}T^2}{2J^2\pi} \nonumber\\
   &+ C
                \Bigr\}, \,\, \calL=\ln\left(
\frac{\sqrt{\pi}e^{\gamma_E+1/4}J}{\sqrt{2}T}
\right). 
\end{align}
Here $\gamma_E$ is the Euler constant, and the integration constant
$C\sim3/4$ is determined by the fitting to the exact result
eq.~\eqref{NLIE}, as in Fig.~\ref{fig:lowT}.

%
%
Note that the leading log in~\eqref{eq:lowHT} 
agrees exactly to what was obtained in
ref.~\cite{OA} via a field theoretical approach.
However, the ``leading log'' is only dominant
in an extremely low temperature and magnetic field.
For a realistic parameter regime, the subleading (in $J/T$ or $J/H$)
corrections determined
in the present study is actually essential in a quantitative discussion
of the shift.

Furthermore, our analysis determines the running coupling
constant with respect to a marginal operator in ref.~\cite{OA} precisely.
Following the logic of ref.~\cite{OA}, this allows a prediction
of the linewidth $\kappa$ including the subleading corrections
for $J'_c\ne0,J'_a=J'_b=0$:
\begin{equation}
\kappa = \frac{\epsilon T}{\pi^3}\left(\frac{J'_c}{J}\right)^2\Bigl\{
               \calL+\frac12\ln\calL+\frac1{4 \calL}+
               \frac{\ln\calL}{4\calL}+C \Bigr\}^2,
\end{equation}
where $\epsilon=4$ ($\epsilon=2$) 
when the applied field $H$ is parallel (perpendicular) to the $c$-axis.
The subleading corrections should be also essential
in a quantitative prediction for realistic values of temperature
and magnetic field.

\begin{figure}[tbh]
\begin{center}
\includegraphics[width=0.45\textwidth]{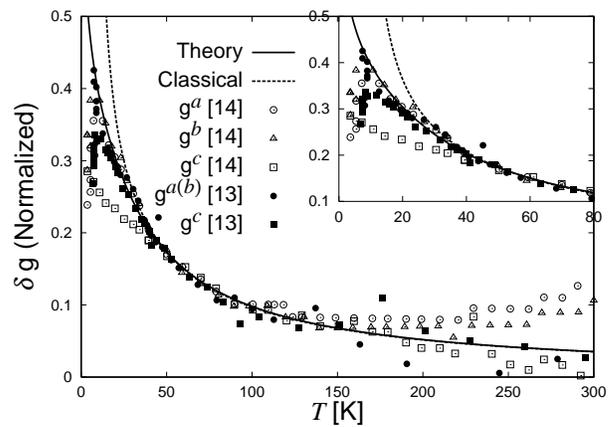}
\end{center}
\caption{
Normalized effective $g$-shift
$\delta  g/(g_\infty f)=(g_{\rm eff}-g_\infty)/(g_\infty f)$
where $f$ is the direction-dependence factor \eqref{eq:f}, 
is plotted against the temperature $T$.
Points are experimental data for LiCuVO$_4$ \cite{exp}
and \cite{Nidda},
while the curve is the theoretical result~\eqref{NLIE}
and the classical results~\eqref{NT}.
Inset: the behavior at low-temperatures.
}
\label{fig:exp}
\end{figure}

Finally, we discuss experimental data in the light of our results.
In Fig.~\ref{fig:exp}, we analyze the
$g$-shift data on LiCuVO$_4$, which
is considered to be an $S=1/2$ Heisenberg 
antiferromagnetic chain. 
The data are taken from ref.~\cite{exp} (powder sample) and
ref.~\cite{Nidda} (single crystal).
We take the estimate $J=44$ [K] from ref.~\cite{exp,Yamaguchi} for the 
intrachain isotropic exchange interaction~\cite{Jnote}. 
Then $H/J=0.027$ and $H/J=0.010$, respectively, for
ref.~\cite{exp} and ref.~\cite{Nidda}.
For these parameters,
in Fig.~\ref{fig:exp} we plot
the theoretical result $Y(T,H)$ from eq.~(\ref{NLIE}).
The difference in the magnetic
field $H$ is negligible in the temperature
regime in which we discuss the experimental data.

The principal axes of the $g$-tensor, i.e. $(g^a,g^b,g^c)$ 
and those of the anisotropic exchange tensor~(\ref{eq:H'}) 
are thought to coincide with the
crystallographic axes $(a,b,c)$~\cite{exp,Nidda}.
Therefore, we use the bare $g$-factor $g^{a,b,c}_\infty$ and the anisotropic
exchange $J'_{a,b,c}$ for each crystallographic axis
as fitting parameters. Here we note that we can set one of
$J'_{a,b,c}$ to zero, by renormalizing the isotropic exchange $J$.
The normalized $g$-shift corresponding to $Y(T,H)$ was then obtained
for each set of experimentally observed effective $g$-factor,
using eqs.~(\ref{eq:f}) and (\ref{R}).

In ref.~\cite{exp}, the anisotropy in $ab$-plane was neglected
in analyzing the powder sample. For this set of data, we obtain
$g^c_\infty=2.22, g^a_\infty = g^b_\infty = 2.09$
as a result of fitting in Fig.~\ref{fig:exp}.
The best fit value of the anisotropic exchange was
$J'_c \sim - 0.15J$ for $H\parallel c$ and $J'_c \sim - 0.11 J$
for $H \perp c$, where we set $J'_a=J'_b=0$.
For the data from ref.~\cite{Nidda}, our fitting yields
$g^c_\infty = 2.31, g^a_\infty = 2.07, g^b_\infty=2.09$.
The set of the anisotropic parameters $J'_c \sim - 0.085J$, $J'_a \sim
9 \times 10^{-4}J$ and $J'_b=0$ well reproduces experimental data 
for the all directions $H\parallel a, b, c$.
These estimates are somewhat different from those in ref.~\cite{Nidda}.
In any case, the observed small anisotropy in the $ab$-plane
is consistent with the treatment in ref.~\cite{exp}.

Overall, the observed temperature dependence is
well reproduced by our theory down close to N\'{e}el temperature
$T_N\sim2.3$ [K], where the interchain interactions should be
important also for ESR.
Discrepancies are however found in the high temperature regime
($> 200$K) for all the three directions of $H$ in ref.~\cite{Nidda},
and in the low temperature regime ($< 50$K) for $H \parallel c$ in
ref.~\cite{Nidda}.
The former could be due to the low signal-to-noise ratio at high temperatures.
The latter data may be also questionable for the absence of the anomaly
(presumably related to the higher dimensional effects)
below 10K, which are observed in all the other data set.

The estimated values of the exchange anisotropy,
especially $J'_c/J \sim - 0.15$ for the data from ref.~\cite{exp},
seems too large compared to the typical value of a few percent for
a Cu ion.
This could be due to a possible overestimation of the $g$-shift
in the analysis in ref.~\cite{exp} to extract
the direction-dependent effective $g$-factor
from the direction-averaged data on the powder sample.

In any case, the temperature dependence of the $g$-shift of LiCuVO$_4$
is better accounted by the present theory than by
the classical one~\cite{NT}, especially at low temperatures.
We hope that the present study stimulates 
further experimental studies on the $S=1/2$ antiferromagnetic chain for
more precise comparisons with the theory.

The authors thank M. Shiroishi and A. Tokuno
for critical reading of the manuscript.
This work was partially supported
by Grant-in-Aid for scientific research
and a 21st Century COE Program at
Tokyo Tech ``Nanometer-Scale Quantum Physics'', both from the
Ministry of Education, Culture, Sports, Science and Technology of Japan.


\begin{thebibliography}{99}
\bibitem{Jnote}
The value of $J$ is twice of $22$K in ref.~\cite{exp,Yamaguchi} due to the
different definition of $J$.
\end{thebibliography}
\end{document}